\documentclass[12pt,onecolumn]{IEEEtran}

\ifCLASSINFOpdf
   \usepackage[pdftex]{graphicx}

\else

\fi

\usepackage{color}
\usepackage[cmex10]{amsmath}
\usepackage{amssymb}   % \triangleq vs. gibi seyler bunda
\usepackage{amsxtra}
\usepackage{amscd}
\usepackage{amsthm}

\usepackage{graphicx}   % sekilller icin

\usepackage{array}  % tablo icin satir acma

\usepackage{multirow}  % tablolar icin coklu satir kullanmada

 \usepackage{flushend}

\usepackage{cite}

\usepackage{color}

\begin{document}
% can use linebreaks \\ within to get better formatting as desire
\title{{\huge Media-Based Modulation for Future Wireless Systems: A Tutorial}}


\author{Ertugrul~Basar,~\IEEEmembership{Senior Member,~IEEE} 
%	\vspace*{-0.35cm}
	\thanks{\textit{E. Basar is with the Communications Research and Innovation Laboratory (\textsc{CoreLab}), Department of Electrical and Electronics Engineering, Ko\c{c} University, Sariyer 34450, Istanbul, Turkey. e-mail: ebasar@ku.edu.tr.}} 
	%This work was supported by the Scientific Research Projects Foundation, Istanbul Technical University.}
	}

\maketitle

\begin{abstract}

The wireless revolution has already started with the specified vision, overall objectives, and the first official 3GPP release of 5th generation (5G) wireless networks. Despite the development of several modern communication technologies, since the beginning of the modern era of digital communications, we have been mostly conveying information by altering the amplitude, the phase, or the frequency of sinusoidal carrier signals, which has inherent drawbacks. On the other hand, index modulation (IM) provides an alternative dimension to transmit digital information: the indices of the corresponding communication systems' building blocks. Media-based modulation (MBM), which is one of the newest and the most prominent members of the IM family, performs the transmission of information by altering the far-field radiation pattern of reconfigurable antennas (RAs) and provides a completely new dimension to convey information: wireless channel fade realizations themselves through the unique signature of received signals. The aim of this article is to shed light on this promising frontier from a broad communication engineering perspective by discussing the most recent advances as well as possible interesting research directions in MBM technologies.

\end{abstract}
\vspace*{-0.1cm}
%\begin{IEEEkeywords}
%OFDM, index modulation, MIMO systems, maximum likelihood (ML) detection, minimum mean square error (MMSE) detection, V-BLAST, 5G wireless networks.
%\end{IEEEkeywords}

% \fi
\IEEEpeerreviewmaketitle

\renewcommand{\thefootnote}{\arabic{footnote}}

\section{Introduction}

%''\textit{The fundamental problem of communication is that of reproducing at one point either exactly or approximately a message selected at another point}''. Although we (communication engineers) have made significant breakthroughs over the past 70 years, this fundamental problem, which is pointed out by Shannon in his landmark 1948 study, is more serious than ever! 

The development of new services/applications, the continuously increasing number of users/devices, and the major advances in technology contribute to the growth of overall data traffic at a high and steady rate. According to the forecasts of Cisco in the global scale, there will be 27.1 billion networked devices and connections in 2021, up from 17.1 billion in 2016, while the internet protocol (IP) traffic will reach an annual run rate of 3.3 Zettabytes in 2021, up from an annual run rate of 1.2 Zettabytes in 2016.

While the above calculations continue to put more and more stress on next-generation communication systems, 5th generation (5G) wireless networks, which are expected to be introduced around 2020, have been under the spotlight over the past few years. Despite the ongoing debate on 5G wireless technology, compared to their fourth generation (4G) counterparts, 5G networks are expected to provide immense bandwidths and much higher data rates with considerably lower latency. 5G wireless networks are also expected to enable a variety of new applications under three main service categories (usage scenarios): i) Enhanced mobile broadband (eMBB), ii) Ultra-reliable and low latency communications (uRLLC), and iii) Massive machine type communications (mMTC), which are identified by the International Telecommunication Union in September 2015 considering the specific needs of diverse applications. 
%While the latest (December 2017) 3GPP Release 15 5G New Radio (NR) specification  mainly focuses on eMBB applications, later Releases are expected to fulfil the requirements of uRLLC and mMTC. 
To meet the goals of 5G NR, the researchers have put forward attractive physical layer (PHY) solutions, such as massive multiple-input multiple-output (MIMO) systems, millimeter-wave (mm-Wave) communications, non-orthogonal multiple access schemes, and novel waveform designs. However, one thing has become certain during the standardization of 5G NR: there is no single enabling technology that can support all of the application requirements being promised by 5G networking. Consequently, to address these different user applications and requirements as well as to support connectivity in the massive scale, 5G and beyond radio access technologies (RATs) should have a strong flexibility to support and employ novel PHY techniques with higher spectral/energy efficiency and lower transceiver complexity. At this point, unconventional transmission methods based on the promising concept of index modulation (IM) \cite{IM_5G}, may have remarkable potential and impact to shape 5G and beyond RATs due to their inherently available advantages over conventional systems. Since the initial skepticism of both academia and industry on the potential and applicability of IM technologies has disappeared now, we believe that IM is not another simple digital modulation alternative, but rather can be a game-changing communication paradigm whose time has come!

Since the beginning of the modern era of digital communications, conventionally, we have been mostly transmitting information by modulating the amplitude, the phase, or the frequency of a high-frequency sinusoidal carrier signal. Consequently, in order to satisfy the increasing demand for higher data rates, we have to employ higher order modulation formats, which require an increased signal power or sophisticated channel encoding methods to achieve a satisfactory error performance. On the other hand, IM provides an alternative way to transmit digital information: the indices of the building blocks of corresponding communication systems, including transmit antennas, subcarriers, time slots, radio frequency (RF) mirrors, and so on \cite{Basar_2017}.

\textit{Media-based modulation (MBM)}, which is one of the newest members of the IM family, can be implemented by intentionally altering the far-field radiation pattern of a reconfigurable antenna (RA) through adjusting the on/off status of its available RF mirrors, which are parasitic RA elements that contain PIN diodes, according to the information bits. In this sense, MBM offers a completely new dimension for the transmission of digital information: the realizations of wireless channels themselves by deliberate perturbation of the transmission environment \cite{Khandani3}. In other words, MBM exploits different realizations of the wireless channels to create its modulation alphabet, that is, it performs the modulation of the wireless channel itself in a sense. Big deal, you may say! Let us consider that you want to transmit six information bits in each signaling interval by using a non-RA (traditional) antenna, which employs the conventional $ 64 $-ary quadrature amplitude modulation ($ 64 $-QAM). In this case, you have to either increase the transmit power or employ sophisticated channel codes to achieve a target bit error rate compared to a low-rate scenario since you inevitably use an over-crowded signal constellation. On the other hand, consider the operation of an RA that has six RF mirrors and transmitting a non-data bearing carrier signal. An RF mirror is a passive antenna element containing a PIN diode, which can be turned on or off according to the information bits. Particular on or off states of the available RF mirrors through the principle of IM can create a different far-field radiation pattern, that is, corresponding to a different realization of the fading channel, which affects the signature of the received signal. For MBM, you do not need to suffer from over-crowded traditional signal constellations and can achieve a much better error performance since the Euclidean distance between MBM constellation points, which are channel fade realizations, remains the same even with increasing data rates. For these reasons, we conclude that MBM, as being an interesting and new wireless communication paradigm, can play a key role in reducing the size, cost, and power consumption of mobile units and Internet of Things (IoT) devices. However, along with accompanying operational difficulties, interesting as well as challenging research problems need to be solved considering the current and relatively premature body of knowledge on MBM technologies. 

In this introductory article, we present the basic concept of MBM, which is still waiting to be explored by many researchers, briefly investigate the most recent advances in this frontier, and outline the possible interesting research directions by discussing the potential of MBM towards next-generation wireless systems. This article, which is the first tutorial-type article in the literature particularly focusing on this frontier, also aims to attract the attention of academia and industry members regarding the promising MBM technologies and provides a broad communication engineering perspective for future MBM systems.

\section{Media-Based Modulation: A Reconfigurable Antenna-Based Index Modulation Scheme}

The development of reconfigurable antennas (RAs), which have the ability to modify their main characteristics including radiation pattern, polarization, or frequency of operation \cite{RA_Survey}, has been one of the most intriguing frontiers in reconfigurable RF research. In most simple terms, an antenna can be reconfigured by altering its radiated fields through deliberate redistribution of antenna currents. RAs can be implemented with different reconfigurable mechanisms, including mechanical modifications, material property changes, and switches. While RAs are generally designed to support different wireless services operating over different frequencies, for embedding information into the antenna states, the use of switches appears as the most attractive solution. 
%These type of RAs has parasitic elements such as RF mirrors, which contain PIN diodes, and these RF mirrors can be turned on or off according to the information bits to alter the far-field radiation pattern of an RA. 
Within this perspective, reconfigurable systems can be regarded as a subgroup of the vast IM family \cite{IM_5G}, in which the indices of parasitic antenna elements are selected with the purpose of information transmission. The flexibility offered by RAs can remarkably reduce the hardware complexity, cost, and size of target systems compared to conventional RF technology, which is mainly based on communication systems with inflexible (non-reconfigurable) hardware components. In this context, RAs can play a key role in the design of next-generation mobile terminals as well as IoT devices and can increase the flexibility of communication systems by exploiting a new degree of freedom for information transfer. It is worth noting that the RA concept is directly related to the antenna itself and different to the concept of smart antennas, which relies on beamforming techniques.

Media-based modulation (MBM) is a term introduced by Khandani in \cite{Khandani1} with the aim of transmitting data using the concept of RAs. Although the RA concept is well-known in the field of electromagnetics and has become increasingly popular in the past decade; MBM, however, has been put forward explicitly to carry additional information through RAs using the concept of IM. The more general channel modulation term has been recently introduced in \cite{STCM} for the family of space-shift keying (SSK) and MBM/RA variations since in all of these schemes, a carrier signal with constant parameters is transmitted while from the perspective of the receiver, different realizations of the wireless channel convey the information. The received signal contains the unique signature of the activated transmit antenna for SSK; however, it contains the unique signature of the selected antenna state (channel realization) for MBM. As a result, MBM appears as an innovative communication phenomenon and comes up with many appealing advantages compared to legacy communication systems, which will be discussed later.

\begin{figure}[t]
	\begin{center}
		{\includegraphics[scale=1]{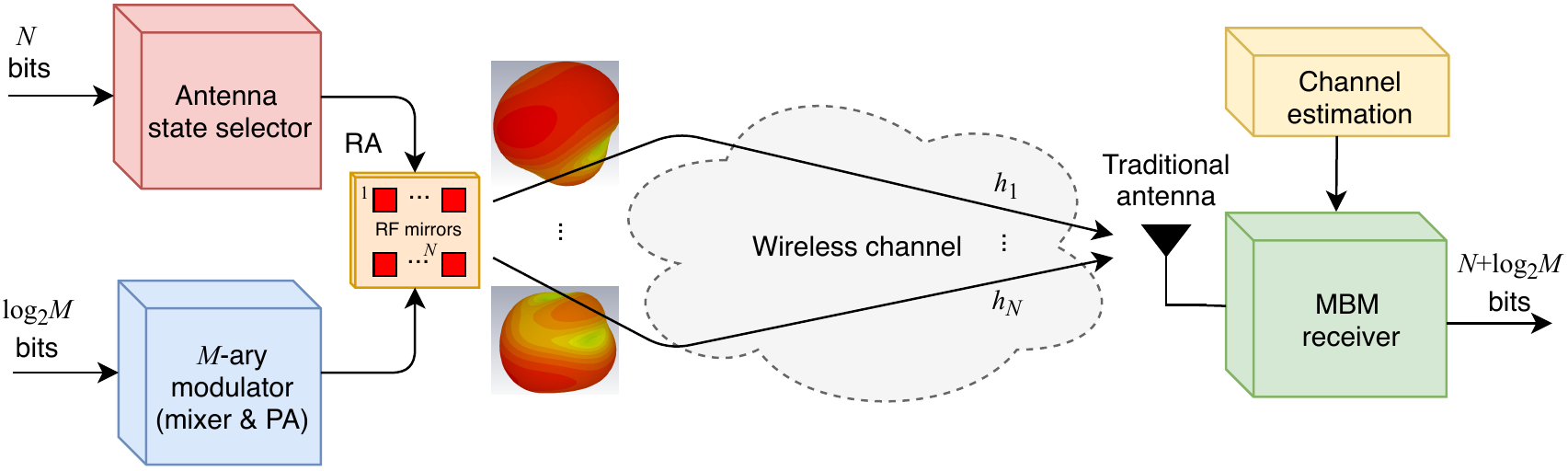}}
		\vspace*{-0.2cm}
		\caption{SISO-MBM transceiver equipped with a transmit RA that contains $N$ RF mirrors.}
		\vspace*{-0.4cm}
	\end{center}
\end{figure}

\subsection{How MBM Works?}

In Fig. 1, the conceptual block diagram of the single-input single-output (SISO)-MBM transceiver is shown, where an RA equipped with $ N $ RF mirrors is considered. In this scenario, the incoming $ N $ bits determine the on/off status (0$\rightarrow$on and 1$\rightarrow$off) of the available $ N $ RF mirrors, whose different on/off combinations correspond to different channel states (realizations) for MBM. Here, $h_1,\ldots,h_N$ represent the corresponding channel fading coefficients (modulation alphabet of MBM) obtained by different channel fade realizations. After I/Q modulation, the data symbol determined by $\log_2 M$ bits is transmitted over this RA. The task of the MBM receiver is to not only recover the $M$-ary modulated data symbol but also estimate the selected channel state (on/off status of the RF mirrors) in order to obtain the bits transmitted by IM. This is accomplished by the maximum likelihood (ML) detector through calculating $2^N  M$ decision metrics. The block diagram of Fig. 1 can be easily generalized for SIMO-MBM, in which each receive antenna experiences $N$ different fade realizations.

The operation of the MBM scheme can be explained by the following example for $N=2$ and $M=4$. In this case, four information bits can be transmitted in a given signaling interval. Let us assume that these four bits are given as $0111$. The first two bits ($01$) of the incoming bit sequence determine the second antenna state for this transmission, while the remaining two bits $(11)$ select the $4$-QAM data symbol of $-1-j$ to be transmitted using the determined antenna state. The receiver, which acquires the knowledge of different fade realizations through an initial training phase, jointly considers all possible antenna states and data symbols (calculating $16$ decision metrics in this example) and determines the most likely pair of antenna state and data symbol.

 Let us consider the case that we want to transmit $ 8 $ bits in each signaling interval by using the schemes of MBM and SSK without considering $ M $-ary modulations. For SSK, we require $ 2^8=256 $ transmit antennas to perform this transmission with antenna indexing, while an MBM scheme with $ 8 $ RF mirrors can handle the same task. Moreover, it has been shown in \cite{STCM} that MIMO-SSK scheme with $T$ transmit antennas is equivalent (in terms of input-output relationship) to the single-input multiple-output (SIMO)-MBM scheme with $ N $ RF mirrors for $T=2^ {N} $.  However, the design of MBM schemes with such high number of RF mirrors (different states) is a challenging research problem as we will discuss later.

%In Fig. 2, we provide our results for a simple computer simulation model of an RA obtained from a dual-band printed G-shaped monopole antenna, which is originally proposed in \cite{Pan_2004} for wireless local area network (WLAN) applications (Fig. 2(a)). 

\begin{figure}[t]
	\begin{center}
		{\includegraphics[scale=0.6]{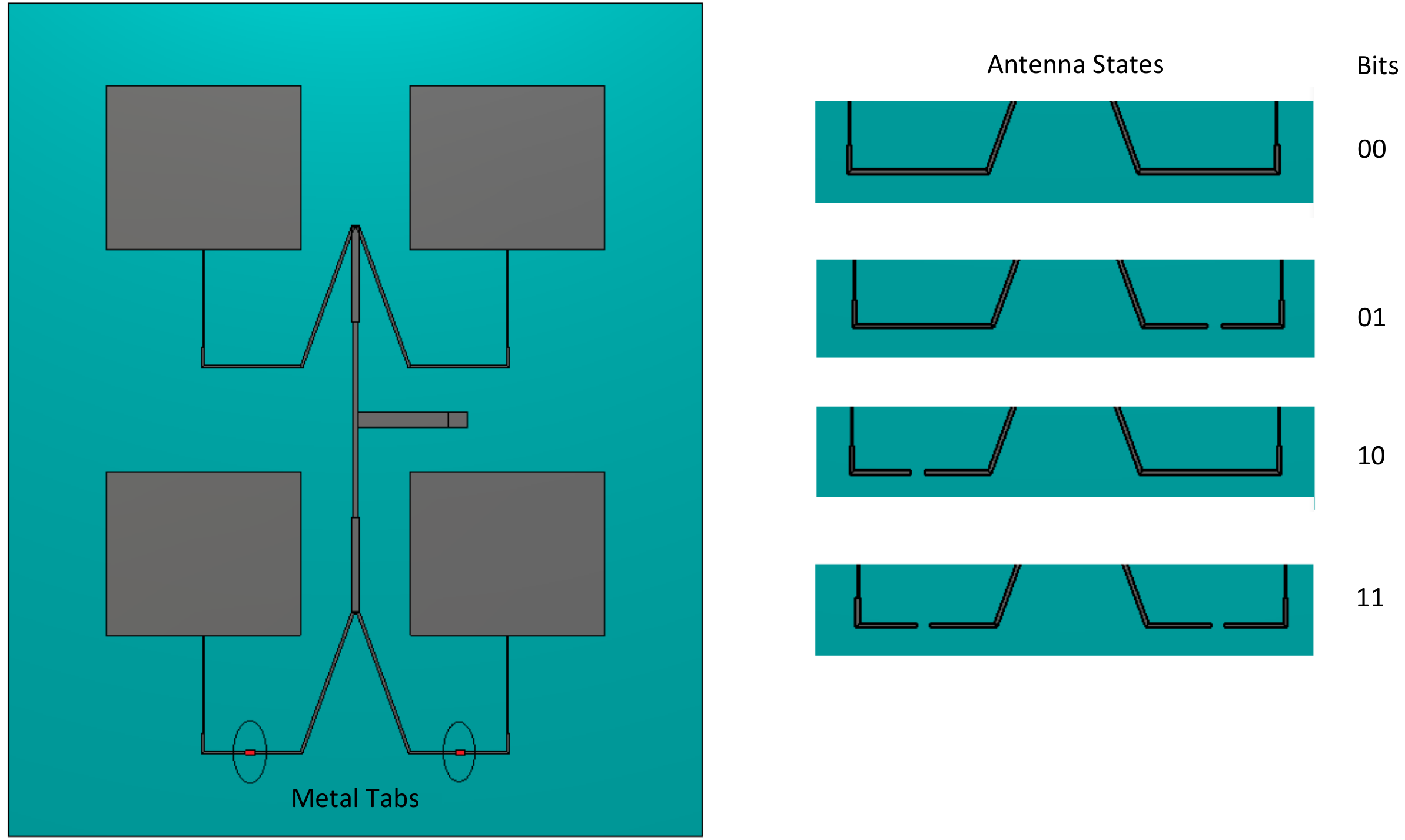}}
		\vspace*{-0.2cm}
		\caption{A simple RA simulation model for MBM, its front view with two ideal metal tabs at lower horizontal connections, and the corresponding four antenna states obtained by altering the status of these two metal tabs.}
		\vspace*{-0.4cm}
	\end{center}
\end{figure}

\subsection{Reconfigurable Antenna Design}
In Fig. 2, we provide an RA simulation model by using a 2-by-2 square patch array consisting of four patch (microstrip) elements. In this model, the effect of PIN diodes (RF mirrors) are modeled by simple (ideal) metal tabs at the lower horizontal connection lines of the array and four different radiation patterns (Fig. 3) or antenna states are obtained by simply including and/or excluding these two metal tabs. The resulting four far-field radiation patterns at $2.4$ GHz are shown in Figs. 3(a)-3(d), where according to the principle of MBM, two additional information bits can be transmitted for this scenario. Specifically, i) For $ 00 $, both metal tabs are included in the design and the radiation pattern of Fig. 3(a) is obtained and used during the signaling interval, ii) For $01$, only the metal tab at the right-hand side is excluded in the design and the radiation pattern of Fig. 3(b) is obtained, iii) For $10$, only the metal tab at the left-hand side is excluded in the design and the radiation pattern of Fig. 3(c) is obtained, and iv) For $11$, both metal tabs are excluded in the design and the radiation pattern of Fig. 3(d) is obtained.
%\begin{itemize}
%	\item For $ 00 $, both metal tabs are included in the design and the radiation pattern of Fig. 3(a) is obtained and used during the signaling interval.
%	\item For $01$, the metal tab at the left-hand side is excluded in the design only and the radiation pattern of Fig. 3(b) is obtained and used during the signaling interval.
%	\item For $10$, the metal tab at the right-hand side is excluded in the design only and the radiation pattern of Fig. 3(c) is obtained and used during the signaling interval.
%	\item For $11$, both metal tabs are excluded in the design and the radiation pattern of Fig. 3(d) is obtained and used during the signaling interval.
%\end{itemize}

\begin{figure}[t]
	\begin{center}
		{\includegraphics[scale=0.5]{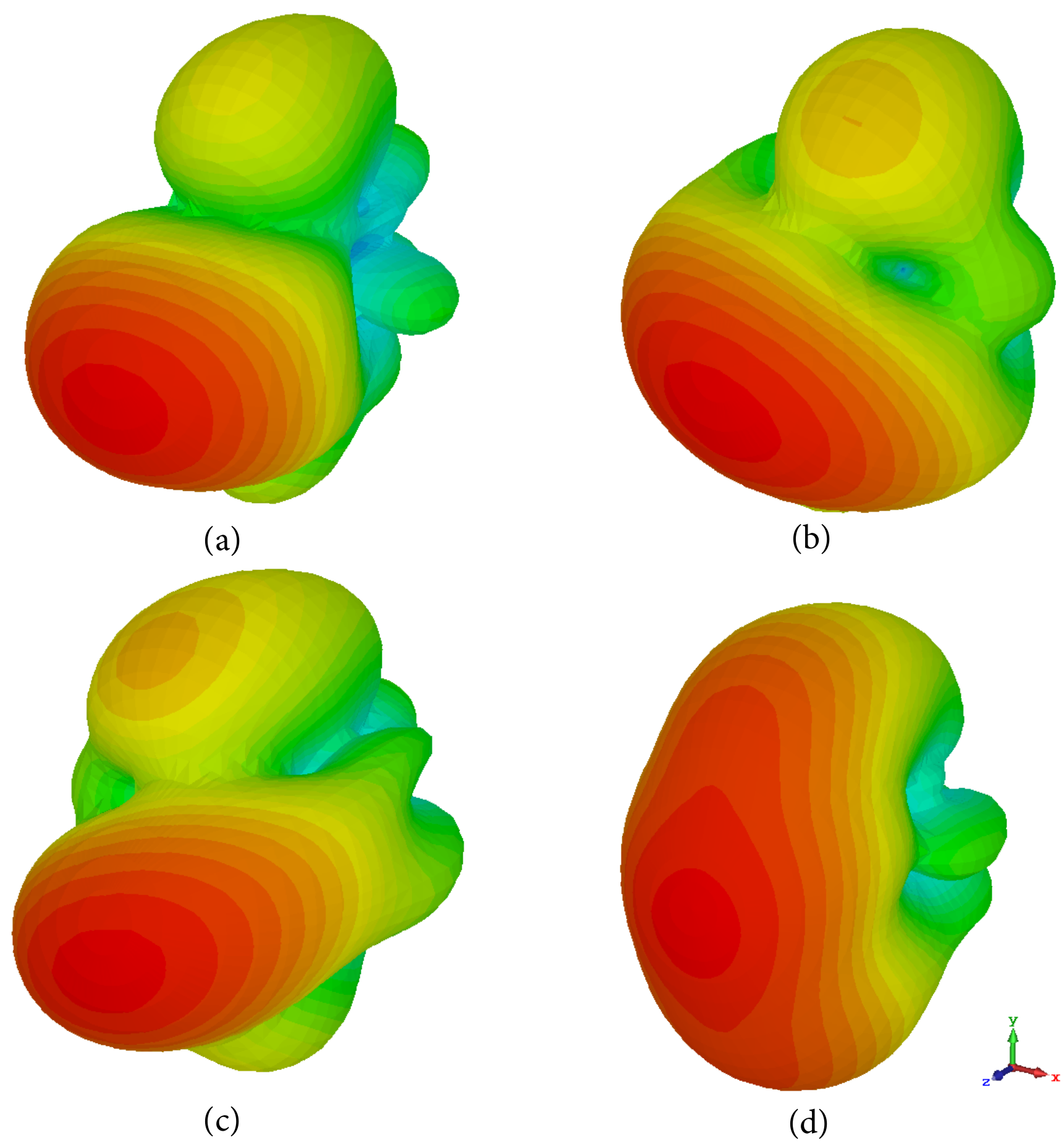}}
		\vspace*{-0.2cm}
		\caption{Generation of four different radiation patterns that can be used in transmission of two bits: (a) State 1, (b) State 2, (c) State 3, (d) State 4.}
		\vspace*{-0.4cm}
	\end{center}
\end{figure}

%For instance, for the information bit of 1, the transmitter employs the radiation pattern of Fig. 2(c), while for the information bit of 0, the pattern of Fig. 2(d) can be employed. 

From the perspective of IM, this scheme can be regarded as pattern shift keying or antenna state shift keying, in which the index of the corresponding radiation pattern is selected according to the information bits. In this context, the more patterns are different, the better is the error performance due to their increasing distinguishability at the receiver. For SISO operation, the modulation alphabet of this generic MBM scheme becomes $\left\lbrace h_1,h_2,h_3,h_4\right\rbrace $ due to the employment of four different antenna states, which correspond to four different fade realizations. It is worth noting that radiation-related parameters, such as reflection coefficient, antenna gain, radiation efficiency, and so on, whose joint optimization is a challenging task and its investigation is beyond the scope of this introductory article, have not been specifically optimized in this scenario, while we simply focus on the modification of the far-field radiation pattern to create the MBM alphabet from the perspective of communication engineering. Nevertheless, the design of Fig. 2 is carefully engineered to ensure that the antenna effectively radiates around 2.4 GHz for all four states. 

For the case of non-line-of-sight (NLoS) transmission, which is commonly modeled by Rayleigh fading, small variations in the radiation pattern of the antenna in a rich scattering environment may result in sufficiently different channel fade realizations due to the perturbation of the propagation environment. In other words, different radiation patterns result in interaction with different scatters, which create different fade realizations at the receiver. While spatial correlation is generally critical for the performance of MIMO systems including SSK, in which the information is embedded into the channel impulse response itself through active antenna selection, the correlation among different radiation patterns (fade realizations) becomes the Achilles' heel of MBM-based systems. A simple correlation model may consider an equal amount of correlation $(\rho)$ between different fade realizations, that is, $E\left\lbrace h_i h_j\right\rbrace =\rho$ for all $i$ and $j$, $i\neq j$, where $E\left\lbrace \cdot \right\rbrace $ is the expectation operator. This model, which is referred to as the equicorrelation model in \cite{MBM_TVT}, can be considered as the worst-case benchmark to model the correlation among different fade realizations while being unrealistic from a practical point of view due to its oversimplified structure.

On the contrary, for LOS transmission with a clear path between the transmitter and the receiver, which is modeled by Rician fading, obtaining sufficiently independent fade realizations for MBM systems becomes a more challenging task. It has been shown that for the case of LOS transmission with a half-wavelength dipole antenna, RAs do not change the envelope fade distribution, while altering the distribution (non-centrality and scaling) parameters of the Rician fading model \cite{RA_SSK}. In other words, RAs may have a noticeable impact on the Rician $K$-factor and the correlation between different fade realizations. However, the work of \cite{RA_SSK} modeled the reconfigurability of the radiation pattern through the inclination angle (mechanical tilt) of the dipole antenna, rendering its practical application cumbersome for high-speed data transmission.

\section{Advantages and Disadvantages}
In this section, we summarize the pros and cons of the MBM scheme and present the trade-offs it offered among complexity, error performance, spectral efficiency, and energy efficiency.

The main advantages of the MBM scheme are summarized as follows:
\begin{itemize}
	\item The scheme of SIMO-MBM is able to create a virtual MIMO system by using only a single RA and can achieve the same error performance as well as spectral efficiency as those of traditional MIMO systems with a considerably more compact transmitter architecture. Furthermore, the SIMO-MBM transmitter requires only a single RF chain, which is one of the most expensive hardware components in the transceiver design with a cost of tens of dollars, and enables a low-cost implementation.
	
	\item Spectral efficiency of MBM increases linearly with the number of parasitic elements (RF mirrors) mounted in RA as long as the independence of the generated radiation patterns is guaranteed.
	%In other words, the spectral efficiency can be doubled by doubling the number of RF mirrors while ensuring a satisfactory performance.
	
	\item Spectral efficiency of the MBM scheme can be significantly boosted by MIMO operation. Existing  MIMO modes such as spatial modulation (SM), generalized SM (GSM), quadrature SM (QSM), and spatial multiplexing (SMUX), can be effectively combined with MBM. In order to improve the diversity order, space-time coding approaches can be integrated to MBM as well. 
	
	\item For the same spectral efficiency, MBM provides a significantly better error performance compared to traditional $ M $-ary modulated systems since the Euclidean distance between MBM constellation points, which are random fade realizations, remains the same even with increasing spectral efficiency values. The advantage of MBM further increases for higher order constellations and a higher number of receive antennas. 
	
	\item To obtain a target error rate, the energy consumption of the MBM transmitter becomes much lower compared to traditional $M$-ary modulated systems, which results in higher energy efficiency. As an example, for a $1\times 8$ SIMO system with $n=8$ bpcu, an uncoded BER value of $10^{-5}$ can be achieved with the MBM scheme at $\sim0$ dB SNR per bit, which is $15$ dB less compared to classical $ 256 $-QAM.
	
	\item The inherent sparsity in the signal model of MBM schemes enables the use of compressed sensing-based detectors that can achieve near-ML performance with a considerably (up to $80-90\%$) lower complexity. Particularly, for increasing spectral efficiency values, the complexity of the brute-force ML detectors becomes enormous, which necessitates efficient low-complexity detection algorithms.
\end{itemize}

In contrast to its appealing advantages listed above, the major drawbacks of the MBM scheme are given as follows:
\begin{itemize}
\item One of the main shortcomings of the MBM scheme is its excessive channel sounding burden. In order to obtain the channel state information (CSI), the receiver has to be trained with pilot signals from all possible antenna states, that is, $2^N$ test signals are required for an RA with $N$ RF mirrors. For fast fading channels, channel estimation at the receiver becomes even more challenging. While this is a major problem for SIMO-MBM, the channel estimation complexity can be significantly reduced by the use of MIMO-MBM with a small number of antenna states at each transmit RA. For the same spectral efficiency, the required number of test signals become $T 2^{N/T}$ for MIMO-MBM, in which $T$ RAs are used, each of them equipped with $N/T$ RF mirrors.

\item MBM comes with challenging design and practical implementation difficulties. First, the design of RAs that can support a high number of sufficiently different radiation patterns is not a straightforward task. In other words, the number of RF mirrors that can be effectively turned on or off may be limited in practice. Furthermore, a compact RA design may limit the total number of RF mirrors that can be accommodated down to one or two. Second, radiation-related parameters have to be carefully monitored to ensure that effective communication is possible with all generated radiation patterns. 	

\item The possible high correlation among different radiation patterns (fade realizations) may become the Achilles' heel of MBM-based systems by limiting the achievable performance. In other words, although a high number of radiation patterns can be generated, a superior performance can be guaranteed only if they can be distinguished from the perspective of IM.
	
\item Similar to all IM-based schemes, the performance of MBM is not satisfactory for a small number of receive antennas, particularly, for a single receiving antenna.

\item Antenna pattern switching between different signaling intervals may cause spectral growth due to the discontinuity in the band-limited MBM signal.	

\end{itemize}

As seen from its advantages and disadvantages summarized above, while MBM provides a new degree of freedom for data transmission and appears as a promising candidate for eMBB as well as IoT applications, it comes with challenging design issues and practical concerns. First, while the increase of the number of antenna states increases the spectral efficiency with a satisfactory performance, this also aggravates the channel estimation burden at the receiver and the complexity of the RA design. Second, although MIMO-MBM provides an easier individual RA design, it requires multiple RF chains when used with the SMUX mode and performs worse than SIMO-MBM in terms of error rate at the same spectral efficiency. Furthermore, MIMO-MBM transmitter may have a higher size due to the employment of multiple RAs and requires highly compact RA designs. Third, the system designer always has to keep an eye on the radiation characteristics as well as the possible correlation among different fade realizations for an increasing number of antenna states and may need to employ  a subset of the available antenna states to obtain a reliable communication system. 

As we will discuss later, interesting research directions can be followed to overcome the aforementioned shortcomings of  MBM schemes.

\subsection{MBM vs Traditional Methods}
In Fig. 4, we compare the bit error rate (BER) performance of classical SIMO $M$-QAM and SIMO-MBM systems for $n=4$ and $n=8$ bits per channel use (bpcu) spectral efficiency values under varying number of receive antennas $R\in\left\lbrace 1,2,4,8,16,32 \right\rbrace $. As seen from Fig. 4, MBM outperforms the classical SIMO scheme for all cases except the case of $R=1$. As mentioned earlier, IM-based schemes generally perform poorly with a single receive antenna. More importantly, the performance gap between MBM and SIMO schemes increases with increasing spectral efficiency.

In Fig. 5, we present the BER performance of classical SIMO $M$-QAM and SIMO-MBM systems for $R=4$ and $R=8$ receive antennas under varying spectral efficiency values $n \in \left\lbrace 1,3,4,8,10,12\right\rbrace $ bpcu. As seen from Fig. 5, while the BER performance of MBM firstly improves and then saturates for $R=8$ with increasing spectral efficiency, which is quite unusual for traditional systems and may have paramount importance for power-limited devices, for the case of $R=4$, BER performance starts to degrade after $n >8$ bpcu. This example proves the superior error performance of MBM with increasing number of receive antennas and spectral efficiency values and the obtained results can be explained by the fact that the pairwise Euclidean distances between MBM constellation points, which are independent fade realizations, do not decrease in any case.

%\section{SIMO-MBM vs MIMO-MBM}

\begin{figure}[!t]
	\begin{center}
		{\includegraphics[scale=0.70]{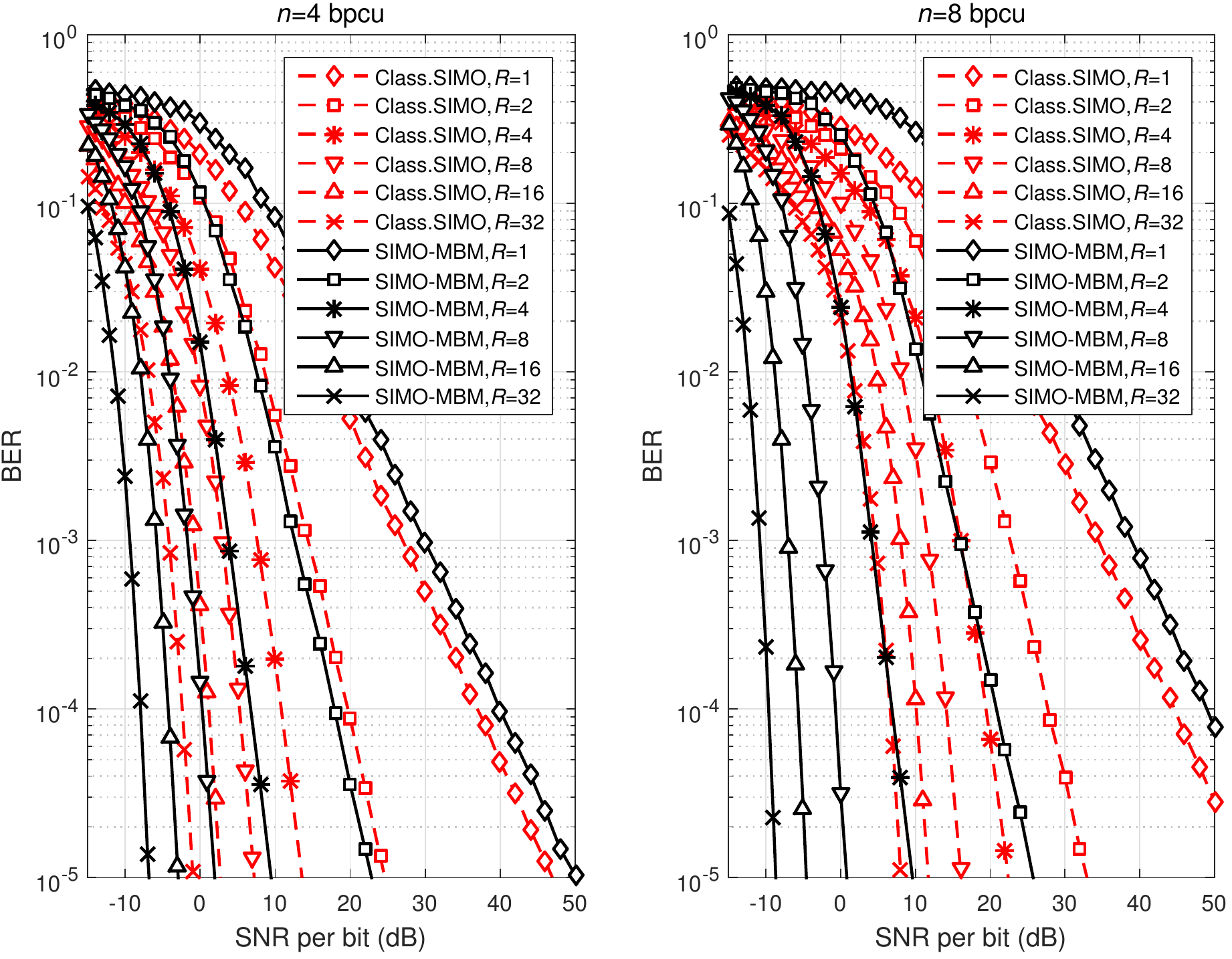}}
		\vspace*{-0.2cm}
		\caption{BER performance comparison of classical SIMO $M$-QAM and SIMO-MBM systems for $n=4$ bpcu (classical SIMO with $16$-QAM and MBM-SIMO with $N=4$) and $n=8$ bpcu (classical SIMO with $256$-QAM and MBM-SIMO with $N=8$) under uncorrelated Rayleigh fading channels.}
		\vspace*{-0.3cm}
	\end{center}
\end{figure}

\begin{figure}[!t]
	\begin{center}
		{\includegraphics[scale=0.70]{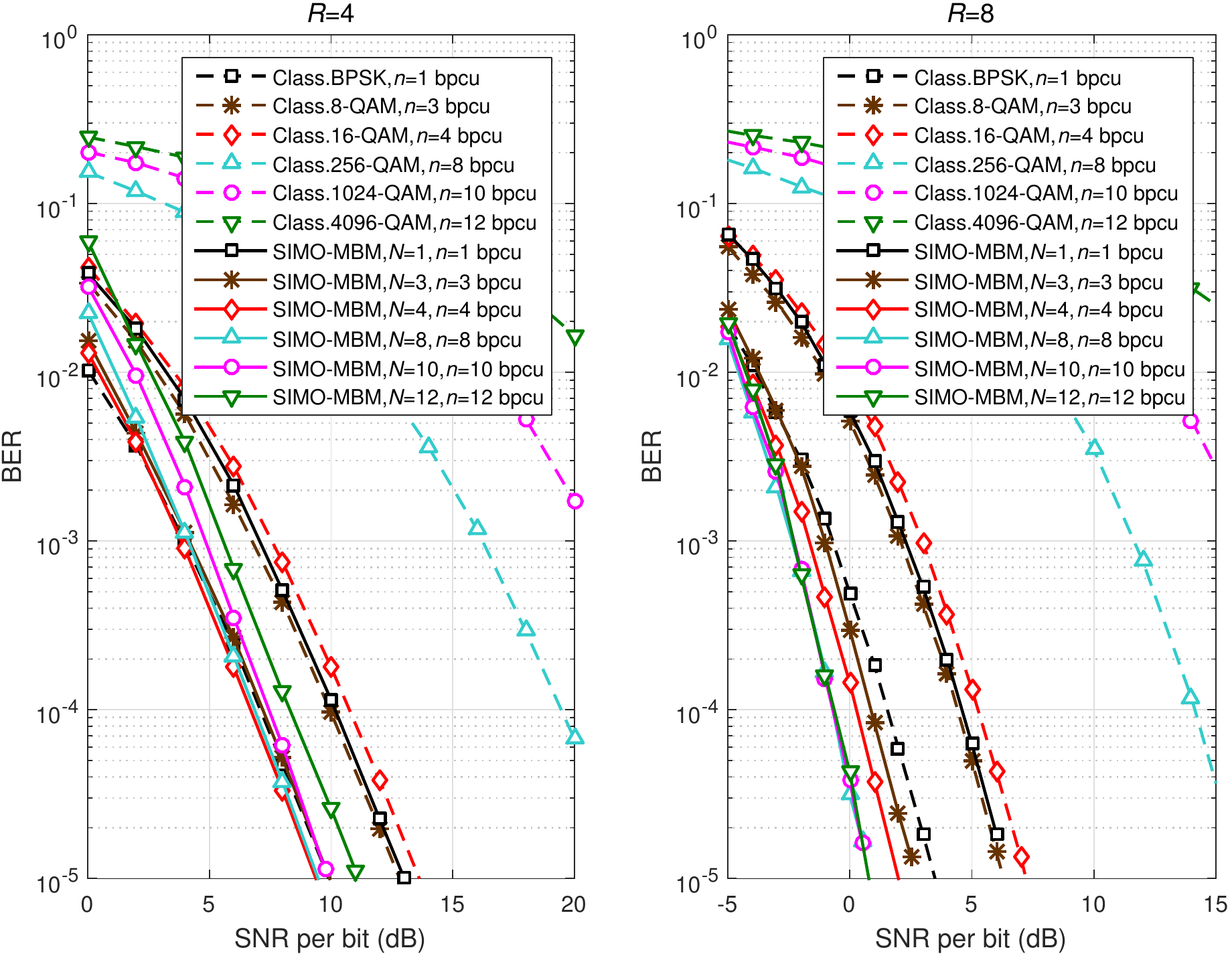}}
		\vspace*{-0.2cm}
		\caption{BER performance comparison of classical SIMO $M$-QAM and SIMO-MBM systems for $R=4$ and $R=8$ receive antennas under uncorrelated Rayleigh fading channels.}
		\vspace*{-0.3cm}
	\end{center}
\end{figure}

\section{State-of-the-Art MBM Solutions}

Although the idea of conveying information through different RA states had been proposed earlier than the introduction of MBM with the aim of improving the performance of phase shift keying schemes, the concept of MBM has triggered a new wave of digital modulation formats to transmit data bits effectively with the deliberate perturbation of the transmission media via adjustable radiation patterns \cite{Khandani1,Khandani3}. 

With the aim of reducing the implementation complexity associated with the transmitter hardware and training overhead, the scheme of SMUX-aided MIMO-MBM is introduced \cite{Khandani3}. The combination of GSM and MBM along with Euclidean distance-based mirror activation pattern selection is later considered to increase the throughput and to obtain additional diversity gains with the feedback of the receiver \cite{MBM_TVT}. MBM is also combined with SSK and QSM, respectively in \cite{RA_SSK} and \cite{QCM}, and promising results reported while ensuring relatively simple MIMO implementation with a single RF chain.

The scheme of dual-polarized SM (DP-SM), in which  one extra information bit is conveyed by the selection of the polarization, is considered in \cite{DP_SM}. Recently, the promising multi-dimensional IM concept is presented by considering the broad applicability of IM techniques and time-indexed MBM, SM-MBM and time-indexed SM-MBM (TI-SM-MBM) schemes are introduced \cite{Multi_IM_Access}. Additionally, load modulation schemes are investigated by modulating the antenna impedances that control the antenna currents. The researchers have also shown the practical applicability of the MBM concept and transmitted information by the realization of four different patterns directed to four different receive antennas \cite{MDR_13}.

In order to overcome the problem of channel sounding, which is a major issue for the MBM receiver with the increasing number of RF mirrors, a differential MBM (DMBM) scheme is proposed \cite{Diff_MBM}. It is worth noting that this scheme does not require CSI knowledge at the receiver and achieves approximately $2$-$4$ dB worse BER performance than the coherent MBM scheme. 

The space-time channel modulation (STCM) scheme, which integrates the space-time block coding concept into MBM with the aim of obtaining transmit diversity gains, is introduced in \cite{STCM}. Recently, the performance of MBM has been investigated in the presence of channel estimation errors \cite{Naresh_2018} and it has been shown that GSM-MBM is considerably robust to channel estimation errors. Even more recently, a low-complexity MBM signal detection algorithm is proposed for large-scale multi-user MIMO systems, in which the multiple users exploit MBM for their uplink transmission \cite{Zhang_2018}.

In Table I, the main MBM schemes investigated in this section are compared in terms of transmitter complexity (cost and size), ML detection complexity, channel estimation complexity, spectral efficiency, and error performance. It should be noted that small variations may exist among different schemes under the same level. As seen from Table I, interesting trade-offs exist among different system metrics with the considered MBM solutions.

\begin{table}[]
	\centering
	\caption{Comparison of Main MBM Schemes}
	\begin{tabular}{cccccc}
		Scheme   & \begin{tabular}[c]{@{}c@{}}Transmitter \\ complexity\end{tabular} & \begin{tabular}[c]{@{}c@{}}ML detection\\ complexity\end{tabular} & \begin{tabular}[c]{@{}c@{}}Channel est.\\ complexity\end{tabular} & \begin{tabular}[c]{@{}c@{}}Spectral \\ efficiency\end{tabular} & \begin{tabular}[c]{@{}c@{}}Error \\ performance\end{tabular} \\ \hline\hline \noalign{\smallskip}
		
		SIMO-MBM \cite{Khandani1} & Low                                                      & Low                                                               & High                                                              & Moderate                                                       & Moderate                                                     \\
		SMUX-MBM \cite{Khandani3} & High                                                     & High                                                              & Moderate                                                          & High                                                           & Low                                                          \\
		SM-MBM \cite{Multi_IM_Access}   & Low                                                      & Low                                                               & Moderate                                                          & Moderate                                                       & Moderate                                                     \\
		SSK-MBM \cite{RA_SSK}  & Low                                                      & Low                                                               & Moderate                                                          & Moderate                                                       & Moderate                                                     \\
		GSM-MBM \cite{MBM_TVT}  & Moderate                                                 & Moderate                                                          & Moderate                                                          & High                                                           & Moderate                                                     \\
		DP-SM \cite{DP_SM}   & Low                                                      & Low                                                               & Low                                                               & Low                                                            & Moderate                                                     \\
		DMBM \cite{Diff_MBM}     & Low                                                      & High                                                              & Low                                                               & Low                                                            & Low                                                          \\
		QCM \cite{QCM}     & Low                                                      & Moderate                                                          & Moderate                                                          & High                                                           & Moderate                                                     \\
		STCM \cite{STCM}   & Moderate                                                 & Moderate                                                          & Moderate                                                          & Moderate                                                       & High          \\ \hline                                               
	\end{tabular}
\end{table}

\section{Unsolved Problems and Possible Research Directions}

The main unsolved problems and possible future research directions in MBM technologies are listed as follows:
\begin{itemize}
	\item For implementation of MBM schemes, novel RA architectures that can generate a sufficiently high number of antenna states with relatively low correlation, have to be designed. Furthermore, the designed RAs have to radiate efficiently for all possible states at the same frequency band and need to be compact in size for possible MIMO employment or IoT applications.
	
	\item Accurate and more realistic mathematical correlation models are needed to quantify the amount of correlation among different fade realizations as well as individual antenna patterns of different RAs in a MIMO configuration. These correlation models have to consider different antenna types, operating frequencies, and propagation environments.
	
	\item Novel SIMO- and MIMO-based MBM transceiver architectures with high spectral efficiency and/or improved error performance can be designed for diverse 5G and beyond application categories. Furthermore, the potential of MBM and its variants can be investigated for both uplink and downlink of massive multi-user systems and relaying networks.
	
	\item Low-complexity ML/near-ML/sub-optimal MBM detectors are required to unlock the potential of MBM schemes at high spectral efficiency values. Adaptive, selection-based, and precoded MBM schemes, which exploit CSI at the transmitter, can be developed as well to further improve the link reliability and throughput.
	
	\item More comprehensive practical implementation campaigns and measurements over practical setups need to be carried out to assess the performance of MBM technologies in real-world scenarios.
\end{itemize}

As seen from above, interdisciplinary research between communication and antenna experts is required to solve some of these open problems.

\section{Conclusions}
While the focus is on increasing the bandwidth by going to mm-Wave bands, using massive MIMO setups to enhance the connectivity as well as considering sophisticated signal processing and channel coding techniques to improve the reliability, MBM emerges as an upcoming technology that can be an alternative and/or supplementary to these modern communication paradigms for beyond 5G networking. This article has provided a new perspective on the potential and applicability of MBM technologies by introducing the basic MBM concept and discussing its major advantages/disadvantages, state-of-the-art MBM solutions, and open research problems. We hope that much more will come shortly in this new and promising communication frontier, stay tuned!    

\section{Acknowledgement}
This work was supported by the Scientific and Technological Research Council of Turkey (TUBITAK) under Grant 117E869. The work of E. Basar was also supported by the Turkish Academy of Sciences GEBIP Programme and the Science Academy BAGEP Programme.

\bibliographystyle{IEEEtran}
\bibliography{IEEEabrv,bib_2018}

\begin{thebibliography}{10}
\providecommand{\url}[1]{#1}
\csname url@samestyle\endcsname
\providecommand{\newblock}{\relax}
\providecommand{\bibinfo}[2]{#2}
\providecommand{\BIBentrySTDinterwordspacing}{\spaceskip=0pt\relax}
\providecommand{\BIBentryALTinterwordstretchfactor}{4}
\providecommand{\BIBentryALTinterwordspacing}{\spaceskip=\fontdimen2\font plus
\BIBentryALTinterwordstretchfactor\fontdimen3\font minus
  \fontdimen4\font\relax}
\providecommand{\BIBforeignlanguage}[2]{{%
\expandafter\ifx\csname l@#1\endcsname\relax
\typeout{** WARNING: IEEEtran.bst: No hyphenation pattern has been}%
\typeout{** loaded for the language `#1'. Using the pattern for}%
\typeout{** the default language instead.}%
\else
\language=\csname l@#1\endcsname
\fi
#2}}
\providecommand{\BIBdecl}{\relax}
\BIBdecl

\bibitem{IM_5G}
E.~Basar, ``Index modulation techniques for 5{G} wireless networks,''
  \emph{IEEE Commun. Mag.}, vol.~54, no.~7, pp. 168--175, June 2016.

\bibitem{Basar_2017}
E.~Basar, M.~Wen, R.~Mesleh, M.~D. Renzo, Y.~Xiao, and H.~Haas, ``Index
  modulation techniques for next-generation wireless networks,'' \emph{IEEE
  Access}, vol.~5, pp. 16\,693--16\,746, Sep. 2017.

\bibitem{Khandani3}
\BIBentryALTinterwordspacing
E.~Seifi, M.~Atamanesh, and A.~K. Khandani, ``Media-based {MIMO}: {A} new
  frontier in wireless communication,'' Oct. 2015. [Online]. Available:
  \url{arxiv.org/abs/1507.07516}
\BIBentrySTDinterwordspacing

\bibitem{RA_Survey}
J.~Costantine, Y.~Tawk, S.~E. Barbin, and C.~G. Christodoulou, ``Reconfigurable
  antennas: {D}esign and applications,'' \emph{Proc. IEEE}, vol. 103, no.~3,
  pp. 424--437, Mar. 2015.

\bibitem{Khandani1}
A.~K. Khandani, ``Media-based modulation: {A} new approach to wireless
  transmission,'' in \emph{Proc. IEEE Int. Symp. Inf. Theory}, Istanbul,
  Turkey, Jul. 2013, pp. 3050--3054.

\bibitem{STCM}
E.~Basar and I.~Altunbas, ``Space-time channel modulation,'' \emph{IEEE Trans.
  Veh. Technol.}, vol.~66, no.~8, pp. 7609--7614, Aug. 2017.

\bibitem{MBM_TVT}
Y.~Naresh and A.~Chockalingam, ``On media-based modulation using {RF}
  mirrors,'' \emph{IEEE Trans. Veh. Technol.}, vol.~66, no.~6, pp. 4967--4983,
  June 2017.

\bibitem{RA_SSK}
Z.~Bouida, H.~El-Sallabi, A.~Ghrayeb, and K.~A. Qaraqe, ``Reconfigurable
  antenna-based space-shift keying ({SSK}) for {MIMO} {R}ician channels,''
  \emph{IEEE Trans. Wireless Commun.}, vol.~15, no.~1, pp. 446--457, Jan. 2016.

\bibitem{QCM}
I.~Yildirim, E.~Basar, and I.~Altunbas, ``Quadrature channel modulation,''
  \emph{IEEE Wireless Commun. Lett.}, vol.~6, no.~6, pp. 790--793, Dec. 2017.

\bibitem{DP_SM}
G.~Zafari, M.~Koca, and H.~Sari, ``Dual-polarized spatial modulation over
  correlated fading channels,'' \emph{IEEE Trans. Commun.}, vol.~65, no.~3, pp.
  1336--1352, Mar. 2017.

\bibitem{Multi_IM_Access}
B.~Shamasundar, S.~Bhat, S.~Jacob, and A.~Chockalingam, ``Multidimensional
  index modulation in wireless communications,'' \emph{IEEE Access}, vol.~6,
  pp. 589--604, Feb. 2018.

\bibitem{MDR_13}
D.-T. Phan-Huy\textit{ et al.}, ``First visual demonstration of transmit and
  receive spatial modulations using the radio wave display,'' in \emph{Proc.
  21st Int. ITG Workshop on Smart Antennas}, Berlin, Germany, Mar. 2017, pp.
  1--7.

\bibitem{Diff_MBM}
Y.~Naresh and A.~Chockalingam, ``A low-complexity maximum-likelihood detector
  for differential media-based modulation,'' \emph{IEEE Commun. Lett.},
  vol.~21, no.~10, pp. 2158--2161, Oct. 2017.

\bibitem{Naresh_2018}
------, ``Performance analysis of media-based modulation with imperfect channel
  state information,'' \emph{IEEE Trans. Veh. Technol.}, vol.~67, no.~5, pp.
  4192--4207, May 2018.

\bibitem{Zhang_2018}
L.~Zhang, M.~Zhao, and L.~Li, ``Low-complexity multi-user detection for {MBM}
  in uplink large-scale {MIMO} systems,'' \emph{IEEE Commun. Lett.}, vol.~22,
  pp. 1568--1571, Aug. 2018.

\end{thebibliography}

%\newpage
\section*{Biography}
Ertugrul Basar (S'09-M'13-SM'16) received the B.S. degree (Hons.) from Istanbul University, Turkey, in 2007, and the M.S. and Ph.D. degrees from Istanbul Technical University, Turkey, in 2009 and 2013, respectively. From 2011 to 2012, he was with the Department of Electrical Engineering, Princeton University, Princeton, NJ, USA, as a Visiting Research Collaborator. He is currently an Associate Professor with the Department of Electrical and Electronics Engineering, Ko\c{c} University, Istanbul, Turkey. Previously, he was an Assistant Professor and an Associate Professor with Istanbul Technical University from 2014 to 2017 and 2017 to 2018, respectively. He is an inventor of four pending/granted patents on index modulation schemes. His primary research interests include MIMO systems, index modulation, cooperative communications, OFDM, visible light communications, and signal processing for communications.

Recent recognition of his research includes the Science Academy (Turkey) Young Scientists (BAGEP) Award in 2018, Turkish Academy of Sciences Outstanding Young Scientist (TUBA-GEBIP) Award in 2017, the first-ever IEEE Turkey Research Encouragement Award in 2017, and the Istanbul Technical University Best Ph.D. Thesis Award in 2014. He is also the recipient of five Best Paper Awards including one from the IEEE International Conference on Communications 2016. He has served as a TPC track chair or a TPC member for several IEEE conferences including GLOBECOM, VTC, PIMRC, and so on. 

Dr. Basar currently serves as an Editor of the \textsc{IEEE Transactions on Communications} and \textit{Physical Communication} (Elsevier), and as an Associate Editor of the \textsc{IEEE Communications Letters}. He served as an Associate Editor for the \textsc{IEEE Access} from 2016 to 2018. He is also the Lead Guest Editor of \textsc{IEEE Journal of Selected Topics in Signal Processing} October 2019 Special Issue ``\textit{Index Modulation for Future Wireless Networks: A Signal Processing Perspective}" and \textit{Physical Communication} January 2019 Special Issue ``\textit{Radio Access Technologies for Beyond 5G Wireless Networks}".

%\section*{Word Count}
%Word Count:4XXX (Introduction through Conclusions, excluding figures, tables, captions, Abstract and References)

\end{document}